\documentclass[12pt]{JHEP3}
\usepackage{amssymb}
\usepackage{bbold}
\usepackage{graphicx}

\title{\bf P-term Potentials from 4-D Supergravity}
\author{C. Burrage\\ E-mail:  \email{C.Burrage@damtp.cam.ac.uk}\\Department of Applied Mathematics and Theoretical Physics\\
\ Centre for Mathematical Sciences\\
 Cambridge CB2 0WA, United Kingdom} 
\author{A.C. Davis\\ E-mail:  \email{A.C.Davis@damtp.cam.ac.uk}\\ Department of Applied Mathematics and Theoretical Physics\\
 Centre for Mathematical Sciences\\
 Cambridge CB2 0WA, United Kingdom} 

\abstract
{P-term inflation arises in some models of brane inflation.  Within
N=2 supersymmetry the scalar potential contains a vector of
Fayet-Iliopoulos (FI)
terms $\xi_i$. Depending on the direction of this vector it is possible to get D-term and F-term inflation or
a mix of these models.  In this paper we review the problems of
embedding the P-term model in supergravity and show how these can be
solved by considering the truncation from an N=2 theory to N=1. We
show that with a simple gauging the scalar potential can include
F-term or D-term parts but not both. The gauging can
be altered so that both F-terms and D-terms containing FI constants
can be included. In all cases we display the inflationary trajectory
and, if it exists, the supersymmetric minimum.}

\begin{document}
\bibliographystyle{JHEP}

\section{Introduction}

P-term inflation is a type of hybrid
inflation \cite{Linde} that appears as the effective theory of some
brane inflation models, in particular the D3/D7 version
\cite{Dasgupta:2004dw,Dasgupta:2002ew} of the KKLMMT
scenario \cite{Kachru:2003aw,Kachru:2003sx}.   In
the framework of N=2 supersymmetric gauge theory a P-term model was
introduced in \cite{Kallosh:2001tm}, with a 
global SU(2,2$|$2) superconformal gauge theory that corresponds to a dual
gauge theory of supersymmetric D3/D7 branes \cite{Kallosh:2003ux}.  Superconformal
SU(2,2$|$2) symmetry can be broken down to N=2 supersymmetry by the vev of
the auxiliary triplet field $P_i$ of the vector multiplet.   The P-term model in N=2 supersymmetry contains a triplet of FI terms $\xi_i$
which arise from a  magnetic flux triplet in the D3/D7 brane
construction. Truncating one of the supersymmetries leaves an N=1 model, which describes both F- and D-term potentials and also potentials which are a mix of the two. 

The status of the
FI terms becomes more complicated when the supersymmetry is made
local.  In \cite{Freedman:1977,Stelle:1978wj}  a supergravity was constructed with a locally
supersymmetric extension of the FI term of the Abelian vector
multiplet.  Local SUSY with an FI term requires axial gauging of the
gravitino and the gaugino, known as local R-symmetry. It has long been known \cite{Barbieri:1982ac}
that a U(1) gauge theory with an FI term can be coupled to
supergravity in a way that preserves the gauge invariance only if the
superpotential of the theory is invariant under the R symmetry and
charged under the U(1) symmetry.   The behaviour
of N=1 supergravity with FI terms and an R-symmetry was
studied in \cite{Binetruy:2004hh} using the superconformal approach
to supergravity \cite{Kallosh:2000ve}.  In the superconformal picture gauge symmetries commute with local
superconformal symmetries, which is shown to
imply that in supergravity a superpotential cannot be gauge invariant.  This has
consequences for the construction of P-term potentials in supergravity
as the charges of
the scalar fields differ in supergravity from their supersymmetric
values, as will be discussed in section \ref{sec:rigid}.  A straightforward generalisation of the
supersymmetric formalism is not possible; to study
P-term inflation it is necessary to construct a P-term potential in supergravity.

  In a supersymmetric P-term model we
start with two supersymmetries and an SO(3) choice of direction for
the FI vector.  One of the symmetries is then broken to leave an N=1 SUSY theory with a
P-term potential.  We wish to repeat this procedure in supergravity; to
construct a P-term model in the N=2 theory and break one of the
symmetries to leave an N=1 P-term
theory.  However until recently it was not known how to obtain FI
terms in N=2 supergravity, embedding the P-term model into supergravity
was not thought to be possible.   In \cite{Achucarro:2005vz} a model was constructed which had FI constants
in the D-term of an N=2 theory. Section \ref{sec:sugra} describes a generalisation of
this construction in which FI terms are also produced in the
superpotential giving rise to a P-term supergravity potential.

The paper is organised as follows.  Section \ref{sec:rigid} reviews the supersymmetric
version of the P-term model in flat and curved spacetime.  A heuristic approach to the
coupling to gravity is taken \cite{Kallosh:2003ux} whereby one of the supersymmetries is made local
and the resulting theory is treated as N=1 supergravity.
The problem of how to embed P-term inflation into
supergravity is addressed in Section \ref{sec:sugra}.  Section \ref{reduction} reviews the way an N=2 supergravity theory is
truncated to an N=1 theory and the conditions this imposes on the
content of the theory.  Section \ref{building} gives the details of
the geometry we use to construct a P-term potential.  We show how a
particular choice of Killing vector for the quaternionic geometry can
give rise to FI terms, however in Section \ref{gauge1} we show that
gauging only one isometry means that D-terms or F-terms are allowed in the reduced potential, but not both.  Sections
\ref{sec:gauging} and \ref{gauge3} show how more complicated gaugings can
give rise to FI terms in the superpotential and D-term of the theory
and we construct the resultant P-term potential.  For each gauging we
consider the vacuum structure of the reduced scalar potential and
discuss its usefulness for cosmology.
We conclude in Section \ref{conc}.

\section{P-term Models in Supersymmetry and Rigid Supergravity}
\label{sec:rigid}
We begin with a review of the P-term potential in supersymmetry.
Take an N=2 supersymmetric theory which contains one hypermultiplet and one
vector multiplet and arrange the two charged scalar fields into a
multiplet of charge $+1$
\begin{equation}
\label{multiplet}
h=\left(\begin{array}{c}
\phi_{+}\\
\phi_{-}^*
\end{array}\right)
\end{equation}
 so that the P-term potential for bosonic scalars can be written
\begin{equation}
V=\frac{g^2}{8}\sum_{i=1}^3(h^{\dagger}\sigma_i h - \xi_i)^2 +
\frac{g^2}{2}|\phi_0|^2 h^{\dagger}h
\label{Spotential} 
\end{equation}
This potential was constructed in \cite{Kallosh:2003ux}. $\phi_0$ is the scalar in the uncharged chiral multiplet, $g$ is a
gauge coupling constant, $\sigma_i$ are the Pauli matrices and $\xi_i$ is a constant Fayet-Iliopoulos
3-vector.  If
$\xi_1 = \xi_2 = 0$
this potential is the super-Bogomol'nyi limit\footnote{as
  defined in \cite{Achucarro:2004ry}} of a D-term potential; if
$\xi_2 = \xi_3 = 0$ this is the Bogomol'nyi limit of an
F-term potential.  We split the three vector $\xi_i$ into the
product of a rotational part described by a SO(3) matrix $R$
and a magnitudinal part $(0,0,\xi)$
\begin{equation}
\label{xii}
\xi_i = (R^{-1})_{ij}\delta_{j3}\xi
\end{equation}
If $R_{ij}=\delta_{ij}$ then (\ref{Spotential}) is a D-term potential, so all P-term
potentials are described by rotations from the D-term case.
 The class of P-term potentials is parameterised by two Euler
 angles, $\psi$, $\theta$, and $\xi$, the magnitude of the FI vector.
 Our conventions for Euler angles are given in appendix \ref{angles}.

  The scalar potential is  constructed from the superpotential and D-term as $V=|\partial W|^2+\frac{g^2}{2}D^2$ with
\begin{equation}
\label{sup}
W=\frac{g\phi_0}{2\sqrt{2}}(P_1-iP_2)
\end{equation}
\begin{equation}
\label{D}
D=\frac{1}{2}P_3
\end{equation}
where
\begin{equation}
\label{P}
P_i = h^{\dagger}\sigma_ih-(R^{-1})_{i3}\xi
\end{equation}
 So a P-term potential contains constant FI terms in the
 superpotential and the D-term.  

To look at inflation in a P-term theory we need to include the
gravitational sector.  One way to proceed
\cite{Kallosh:2003ux} is to couple the N=2
supersymmetric theory to N=1 supergravity by choosing a
supersymmetry, making it local, and treating it  as if it were the only
one. As discussed in the introduction if a  U(1) gauge theory with an FI
term is to be consistently coupled to supergravity in a way that preserves gauge
invariance the superpotential must be invariant under the R-symmetry
but transform under the U(1) symmetry.  This alters the charges of the fields appearing in the superpotential.  In
supersymmetry the scalar fields $\phi_{\pm}$, $\phi_0$ have
charges $Q_{\pm}=\pm1$, $Q_0=0$, and the superpotential
is uncharged.  As an example of how this changes in supergravity take
$W\propto \phi_0 \phi_+ \phi_-$ and an FI constant in the D-term.  Then the fields have charges
\begin{eqnarray}
q_i=Q_i-\rho_i\frac{\xi}{M_P^2} &, & \sum \rho_i=1
\end{eqnarray}
As $M_P\rightarrow \infty$ we regain $q_i=Q_i$, indeed with a generic choice of superpotential the charges always return
to their supersymmetric values in this limit.  $M_P \rightarrow \infty$ is known as the rigid limit of supergravity
\cite{Binetruy:2004hh} and describes supersymmetry in a curved space-time.  Unless the rigid limit is being considered it no longer makes sense to combine $\phi_+$,
$\phi^*_-$ into the multiplet (\ref{multiplet}), so except in this limit
it is unclear how to proceed with this formalism in supergravity.
behaviour in the rigid limit. 

\subsection{Cosmology}
\label{rigidinflate}
To find the form of the scalar potential in rigid supergravity take the minimal K\"{a}hler potential
\begin{eqnarray}
\label{mink}
K&=&\frac{|\phi_+|^2 +|\phi_-|^2 + |\phi_0|^2}{M_P^2}
\end{eqnarray}
minimal vector kinetic terms, and the superpotential (\ref{sup}), and
D-term (\ref{D}). Note that the $M_P \rightarrow \infty$ limit should
not be taken at this stage but at the end of the calculation.  Compute the scalar potential in the usual way
\begin{eqnarray}
\label{formofpot}
V&=&e^K\left\{\left|\frac{\partial W}{\partial
  \phi_i}+\frac{\phi_i^*W}{M_P^2}\right|^2-3\left|\frac{W}{M_P}\right|^2\right\}+\frac{g^2}{2}D^2\label{potexpress}\\
&=&\frac{g^2}{2}\Bigl\{|\phi_+\phi_-|^2+|\phi_0\phi_-|^2+|\phi_0\phi_+|^2-\sin\theta\sin\psi\xi(\phi_+\phi_-+\bar{\phi}_+\bar{\phi}_-)\Bigr.\nonumber\\
& &\;\;\;\;\;\;\;+\Bigl.(\sin\theta\sin\psi\xi)^2\Bigr\} +\mathcal{O}\left(\frac{|\phi_i|^2}{M_P^2}\right) \nonumber\\
& & +\frac{g^2}{8}\Bigl(|\phi_+|^2-|\phi_-|^2-\cos\theta\xi\Bigr)^2
\end{eqnarray}
The term in braces in (\ref{potexpress}) is the F-term part of
the potential, the second term is the D-term.  The full expression for
the potential to all orders in $|\phi_i|/M_P$ is given in appendix
\ref{fulleffpot}.  
This potential has two types of extrema:
\begin{description}

\item[Inflationary Valley.]
$\phi_+=\phi_-=0$ is a stationary point of the potential in the
  $\phi_{\pm}$ directions.  The extremum is a minimum if
\begin{equation} 
2|\phi_0|^2>\xi
\end{equation}
 and a maximum
otherwise.  Inflation occurs when the fields are rolling in
this valley and ends when the critical point is reached.

The effective potential during inflation,  when $\phi_+=\phi_-=0$, is
\begin{equation}
V = \frac{\xi^2g^2}{2}\left\{1
+\sin^2\theta\left[\frac{1}{2}\frac{|\phi_0|^4}{M_p^4} +\mathcal{O}\left(\frac{|\phi_0|^5}{M_P^5}\right)\right]\right\}
\label{effpotential}
\end{equation}
P-term inflationary potentials are parameterised by
$0\leq\sin^2\theta\leq 1$.  The two limiting cases are $\sin\theta=0$
which gives the D-term potential and $\sin^2\theta=1$ which gives the F-term
potential.  In the rigid limit the inflationary potential is flat for all
P-term potentials.  In the D-term case
all corrections to the flat potential vanish which
suggests that the D-term inflationary potential remains flat in supergravity
unlike all other P-term potentials.

\item[Supersymmetric minimum.]  

The
supersymmetric minimum is the point at which 
\begin{equation}
\frac{\partial W}{\partial \phi_i} +\frac{\phi_i^*W}{M_P^2}=0
 \end{equation}
\begin{equation}
D=0
\end{equation}
  This occurs at 
\begin{equation}
\label{susymin}
\begin{array}{c}
\phi_0=0, \\
|\phi_+|^2+|\phi_-|^2=\xi,\\
e^{i\psi}\cos\left(\frac{\theta}{2}\right)\phi_-=i\sin\left(\frac{\theta}{2}\right)\bar{\phi}_+
\end{array}
\end{equation}
where the vacuum manifold is a circle. After the phase transition at the end of inflation the charged fields
waterfall into the supersymmetric minimum.  Cosmic strings may be formed by the Kibble mechanism
\cite{Kibble:1976sj}.

\end{description}
This is the vacuum structure required for hybrid inflation.  All of the P-term models described here give rise to hybrid inflation and allow cosmic string formation at the end of inflation.

\section{P-term models in Supergravity}
\label{sec:sugra}
The analysis of section \ref{sec:rigid}, which describes supersymmetry
in a curved space-time gives an
indication of the behaviour of a P-term model in supergravity.  However
for a complete analysis we need to construct a P-term potential in
supergravity.  This means finding an N=2 supergravity which can be truncated to N=1 in such a way as to give FI terms in the superpotential and D-term of the theory. As in the supersymmetric version of the theory we start with an N=2 supergravity theory containing a constant FI term in each of the three components of the moment map and an arbitrary choice of direction.  We then truncate one of the  supersymmetries to get an N=1 supergravity
theory containing FI terms.  For
notational convenience from now on we will use units in which $M_P=1$.

Consider an N=2 supergravity theory \cite{Andrianopoli:1996vr} with $n_H$ hyper-multiplets, $n_V$ vector multiplets,
 and $n_C$ chiral multiplets.  The hyper-multiplet sector
describes a quaternionic-K\"{a}hler manifold which has SU(2)
connection $\omega^x$, and complex structures $J^x_{uv}$,
 where $x=1,2,3$ and  $u=1,\ldots,4n_H$.  The vector multiplet sector
describes a special-K\"{a}hler Manifold with a $2(n_V+1)$ component
symplectic section
\begin{equation}
V=\left(\begin{array}{c}
L^{\mathbf{\Lambda}}\\
M_{\mathbf{\Lambda}}
\end{array}\right)
=e^{\frac{K}{2}}\left(\begin{array}{c}
X^{\mathbf{\Lambda}}\\
F_{\mathbf{\Lambda}}
\end{array}\right)
\end{equation}
$\mathbf{\Lambda}=0,1,\ldots,n_V$.  $K$ is the K\"{a}hler potential
\begin{equation}
\label{kahler}
e^{-K(z,\bar{z})}=-iX^{\mathbf{\Lambda}}\bar{F}_{\mathbf{\Lambda}}+iF_{\mathbf{\Lambda}}\bar{X}^{\mathbf{\Lambda}}
\end{equation}
where $z^i$ are the coordinates on
the manifold. The symplectic section may be written in terms of a prepotential
$F(X^{\mathbf{\Lambda}})$ so that
$F_{\mathbf{\Lambda}}=\partial F/\partial X^{\mathbf{\Lambda}}$. 

Historically the problem with lifting P-term inflation to supergravity was that it
was not known how to get constant FI terms in N=2 supergravity.  In
N=2 supergravity the moment map which corresponds to the Killing vector
$k_{\mathbf{\Lambda}}$ is given by
\begin{equation}
\label{noPinN2}
4n_H P^x_{\mathbf{\Lambda}} = -J^{xuv}\partial_u k_{v\mathbf{\Lambda}}
\end{equation}
 Constant terms
in the moment maps of the N=2 theory give rise to
FI terms.  Except in a few extremal
cases (for example $n_H=0$, or the rigid limit when the SU(2)
curvature vanishes) it was not known how to add an arbitrary constant
to $P_{\mathbf{\Lambda}}$ and still satisfy (\ref{noPinN2}).  Recently in \cite{Achucarro:2005vz} a moment
map was constructed in N=2 supergravity which gave a D-term potential containing a
constant FI term.  This was done by considering the truncation to an
N=1 theory.  (In \cite{Esole:2006wu} this was extended to admit
an axion-dilaton field).   The moment maps were written as \cite{Galicki:1986ja}
\begin{equation}
\label{compensator}
P^x_{\mathbf{\Lambda}}=\frac{1}{2}r^x_{\mathbf{\Lambda}}+\iota_{\mathbf{\Lambda}}\omega^x
\end{equation}
where $r^x_{\mathbf{\Lambda}}$ is the SU(2) compensator.  A Killing vector
only preserves $\omega^x$ and $J^x$ up to an SU(2) transformation,
and the compensator is defined so that the following equations are satisfied
\begin{equation}
\mathcal{L}_{\mathbf{\Lambda}}\omega^x=\frac{-1}{2}\nabla r^x_{\mathbf{\Lambda}}
\end{equation}
\begin{equation}
\mathcal{L}_{\mathbf{\Lambda}}J^x=\epsilon^{xyz}r^y_{\mathbf{\Lambda}}J^z
\end{equation}
where $\mathcal{L}_{\mathbf{\Lambda}}$ denotes the Lie derivative with
respect to $k_{\mathbf{\Lambda}}$.  In some manifolds it is possible
to choose a Killing vector so that the compensator is constant,
resulting in a 
constant term in the moment map.  In section \ref{building} we show how a particular choice of Killing vectors can
lead to constant
terms in the superpotential and the D-term of the reduced theory and
thus to a P-term potential.

\subsection{Reduction from N=2 to N=1 Supergravity}
\label{reduction}
The relationship
between different P-term models is a result of the symmetries of the
N=2 theory so we begin our construction of a supergravity P-term model
in N=2 supergravity. The only known way to get FI terms in N=2 supergravity is
by truncating the theory to one which resembles an N=1 supergravity
containing FI terms \cite{Achucarro:2005vz} where we can construct a
useful model of inflation.  This parallels the
discussion of P-term inflation in \cite{Kallosh:2003ux} where the P-term model
was constructed by taking an N=2 supersymmetry and then making one of the symmetries
local and disregarding the other to move to N=1 supergravity. 

The
constraints imposed by the truncation from N=2 to N=1 supergravity  are
known \cite{Andrianopoli:2001gm,Andrianopoli:2001zh} and we review them here.  The
truncation is done by constraining the theory so that the N=2
supergravity transformations look like those of an N=1 theory.
Additional conditions come from ensuring that the reduced theory is
consistent.  A full derivation of the truncation conditions is not necessary for this
work;  we state here the conditions which constrain the moment maps or the
Killing vectors of the N=2 theory.  This will allow us to compute the
reduced scalar potential.

The reduction truncates the spin
3/2 multiplet as the second gravitino, $\psi_{2\mu}$, and the
graviphoton are not present in N=1 supergravity.  The N=2 gravitino transformations are 
\begin{equation}
\label{gravitino}
\delta \psi_{A\mu}=\hat{\nabla}_{\mu}\epsilon_A+(igS_{AB}\eta_{\mu
  \nu}+\epsilon_{AB}T^-_{\mu \nu})\gamma^{\nu}\epsilon^B
\end{equation}
where
\begin{equation}
\hat{\nabla}_{\mu}\epsilon_A=\mathcal{D}_{\mu}\epsilon_A+\hat{\omega}_{\mu|A}^B\epsilon_B+\hat{\mathcal{Q}}_{\mu}\epsilon_A
\end{equation}
The SU(2) gauge connection is 
\begin{equation}
 \hat{\omega}_A^B=\omega_A^B+g_{(\mathbf{\Lambda})}A^{\mathbf{\Lambda}}P^x_{\mathbf{\Lambda}}(\sigma^x)_A^B
\end{equation}
and the U(1) gauge  connection is
\begin{equation}
\hat{\mathcal{Q}}=\mathcal{Q} +g_{(\mathbf{\Lambda})}A^{\mathbf{\Lambda}}P^0_{\mathbf{\lambda}}
\end{equation}
where $\omega_A^B$ and $\mathcal{Q}$ are the connections of the
ungauged theory. $P^0_{\mathbf{\Lambda}}$ is the holomorphic
momentum map, the index $0$ is to distinguish it from the
triholomorphic momentum map $P^x_{\mathbf{\Lambda}}$.
 $S_{AB}$ describes
the shift of the gravitino  
\begin{equation}
S_{AB}\equiv\frac{i}{2}P^x_{\mathbf{\Lambda}}\sigma^x_{AB}L^{\mathbf{\Lambda}}
\end{equation}
  The `dressed' graviphoton is
\begin{equation}
T^-_{\mu
  \nu}\equiv2i\mbox{Im}\mathcal{N}_{\mathbf{\Lambda}\mathbf{\Sigma}}L^{\mathbf{\Sigma}}F^{\mathbf{\Lambda}-}_{\mu
  \nu}
\end{equation}
where $\mathcal{N}_{\mathbf{\Lambda}\mathbf{\Sigma}}$ is the N=2 vector
kinetic matrix which can be computed from the prepotential
$F(X^{\mathbf{\Lambda}})$ as
\begin{equation}
\label{veckin}
\mathcal{N}_{\mathbf{\Lambda}\mathbf{\Sigma}}=\bar{F}_{\mathbf{\Lambda}\mathbf{\Sigma}}+\frac{iN_{\mathbf{\Lambda}\mathbf{\Delta}}N_{\mathbf{\Sigma}\mathbf{\Gamma}}X^{\mathbf{\Delta}}X^{\mathbf{\Gamma}}}{N_{\mathbf{\Theta}\mathbf{\Upsilon}}X^{\mathbf{\Theta}}X^{\mathbf{\Upsilon}}}
\end{equation}
with $F_{\mathbf{\Lambda}\mathbf{\Sigma}}=\partial^2F(X^{\mathbf{\Delta}})/\partial X^{\mathbf{\Lambda}}\partial X^{\mathbf{\Sigma}}$ and
$N_{\mathbf{\Lambda}\mathbf{\Sigma}}=2\mbox{Im}F_{\mathbf{\Lambda}\mathbf{\Sigma}}$.

  For the truncation to an N=1
theory  to be consistent the transformation of the second gravitino, $\delta\psi_{2\mu}$, must be set to zero.  In the ungauged
case this means
\begin{equation}
\label{graviphoton}
T^-_{\mu
  \nu}=0
\end{equation}
 The most general way to find solutions to
(\ref{graviphoton}) is to view it as an orthogonality relation between the vector
and scalar sectors.  To satisfy this relation we split the index $\mathbf{\Lambda}=\{\Lambda,X\}$ so that
$\Lambda=1,\ldots,n_V^{\prime}=n_V-n_C$ and runs over the retained
vectors of the reduced
theory and
$X=0,\ldots,n_C$ and runs over the retained scalars.  To satisfy
(\ref{graviphoton}) set $F^X_{\mu\nu}=0$ and
$\mbox{Im}\mathcal{N}_{\Lambda\mathbf{\Sigma}}L^{\mathbf{\Sigma}}=0$
which requires
\begin{equation}
\label{Liszero}
L^{\Lambda}=0
\end{equation}
In a gauged theory the vanishing of the second gravitino gives the
additional constraints $S_{21}=0$ and $\hat{\omega}_1^2=0$.
These are satisfied if
\begin{equation}
\label{Dtermconditions}
P^1_{\Lambda}=P^2_{\Lambda}=0
\end{equation}

In an N=1 theory the gravitino transformation is
\begin{equation}
\delta\psi_{\mu}=D_{\mu}\epsilon+\hat{Q}_{\mu}\epsilon+iL(z,\bar{z})\gamma_{\mu}\epsilon
\end{equation}
where
\begin{equation}
\label{LWK}
L(z,\bar{z})=W(z)e^{\frac{1}{2}K(z,\bar{z})}
\end{equation}
comparing this with the reduction of (\ref{gravitino}) shows that
\begin{eqnarray}
\label{Lequation}
L(z,\bar{z})&=&\frac{i}{2}g_{(\mathbf{\Lambda})}L^{\mathbf{\Lambda}}(P^1_{\mathbf{\Lambda}}-iP^2_{\mathbf{\Lambda}})\nonumber\\
&=&\frac{i}{2}g_{(X)}L^X(P^1_X-iP^2_X)\label{super}
\end{eqnarray}
by (\ref{Liszero}).  

  The reduction of the transformations in the hypermultiplet
and vector multiplet sectors should be considered separately. For the truncated theory to have the correct form for N=1 supergravity the quaternionic manifold
of the hyper-multiplet sector must be reduced to a K\"{a}hler-Hodge one and the
special-K\"{a}hler manifold of the vector sector reduced to a smaller K\"{a}hler manifold.  Reducing the vector multiplet sector requires the truncation of the
gaugino transformations.  The N=1 gaugino transformation is
\begin{equation}
\delta\lambda^{\Lambda}=\mathcal{F}^{-\Lambda}_{\mu\nu}\gamma^{\mu\nu}\epsilon
+iD^{\Lambda}\epsilon
\end{equation}
For this to be the reduction of the N=2 theory
\begin{equation}
\label{DwithP0}
D^{\Lambda}\equiv
(\mbox{Im}\mathcal{N}^{-1})^{\Lambda\Sigma}(P^0_{\Sigma}+P^3_{\Sigma})
\end{equation}
where $P^0_{\Sigma}$ is given by 
\begin{equation}
P^0_{\Sigma}=2i\mbox{Im}\mathcal{N}_{\Sigma\Lambda} f^{\Lambda}_{XY}\bar{L}^XL^Y
\end{equation}
and gives the special-K\"{a}hler
manifold contribution to the D-term.  $f^{\mathbf{\Lambda}}_{\mathbf{\Delta\Sigma}}$ are the structure constants of the
N=2 gauge group.  $P^0_{\mathbf{\Lambda}}$ vanishes for an Abelian gauging.  For
the truncation of the gaugino to be consistent we must ensure
\begin{equation}
\label{superconditions}
P^3_X=0, \;\;\;\;\;\; P^0_X=0
\end{equation}
The truncation also requires that the vector kinetic matrix, $f_{\Lambda\Sigma}$, of the reduced theory is 
\begin{equation}
f_{\Lambda\Sigma}=2\bar{\mathcal{N}}_{\Lambda\Sigma}
\end{equation}

Truncation of the hypermultiplet sector requires the holonomy of the
quaternionic manifold to be reduced because in N=1
supergravity all the scalars must lie
in chiral multiplets with K\"{a}hler-Hodge (K-H) structure.  This
is equivalent to selecting an $n_H$ dimensional complex submanifold on
which the $2n_H$ extra degrees of freedom are frozen.  Let $q^u$ be the coordinates on the
quaternionic manifold, the reduction requires $q^{4s+3}= q^{4s+4}=0$,
$s=0,\ldots,n_H-1$. 
Then $w^s=q^{1+4s}+iq^{2+4s}$  are the $n_H$ holomorphic
coordinates on the K-H manifold.  We define $n^t$ as the
real coordinates along directions orthogonal to the reduced K-H manifold.
This decomposition of indices allows us to write down the last set of
conditions needed for the truncation; by considering the gauge groups in the original
and truncated theories it can be shown that the Killing vectors of
the quaternionic manifold are required to satisfy
 \begin{equation}
\begin{array}{cc}
\label{killingconditions}
k^s_X=0, & k^t_{\Lambda}= 0
\end{array}
\end{equation}
For clarity we will list here all the truncation conditions needed to
construct the scalar potential
\begin{eqnarray}
L^{\Lambda}&=&0\label{Lcon}\\
P^1_{\Lambda}=P^2_{\Lambda}&=&0 \label{conDsup}\\
P^3_X=P^0_X&=&0\\
k^s_X= k^t_{\Lambda}&=&0\label{conkill}
\end{eqnarray}
Note that the
moment maps that generate the superpotential have a scalar index and the
moment maps that generate the D-term have a vector index.  

To compute the P-term potential we start with the N=2 scalar potential
\begin{eqnarray}
V_{\mbox{\tiny N=2}}&=&\bar{L}^{\mathbf{\Lambda}}L^{\mathbf{\Sigma}}\Bigl(g_{\mathcal{I}\bar{\mathcal{J}}}k^{\mathcal{I}}_{\mathbf{\Lambda}}k^{\bar{\mathcal{J}}}_{\mathbf{\Sigma}}+4h_{uv}k^u_{\mathbf{\Lambda}}k^v_
{\mathbf{\Sigma}}\Bigr)\nonumber\\
&
&+P^x_{\mathbf{\Lambda}}P^x_{\mathbf{\Sigma}}\left(\frac{-1}{2}(\mbox{Im}\mathcal{N}^{-1})^{\mathbf{\Lambda}\mathbf{\Sigma}}-\bar{L}^{\mathbf{\Lambda}}L^{\mathbf{\Sigma}}\right)\nonumber\\
& &-3P^x_{\mathbf{\Lambda}}P^x_{\mathbf{\Sigma}}\bar{L}^{\mathbf{\Lambda}}L^{\mathbf{\Sigma}}
\end{eqnarray}
where $h_{uv}$ is the metric on the quaternionic manifold, and $\mathcal{I}=1,\ldots,n_V$ is the world index.
After truncation this becomes
\begin{equation}
\label{potential}
V_{\mbox{\tiny N=1}}=4\left(-3L\bar{L}+g^{i\bar{j}}\nabla_iL\nabla_{\bar{j}}\bar{L}+g^{s\bar{s}}\nabla_sL\nabla_{\bar{s}}\bar{L}
+\frac{1}{16}\mbox{Im}f_{\Lambda \Sigma}D^{\Lambda}D^{\Sigma}\right)
\end{equation}
which can be computed from (\ref{veckin}), (\ref{Lequation}) and (\ref{DwithP0})
 if the moment maps and prepotentials are known.

\subsection{Building a P-term Potential}
\label{building}
We choose the simplest geometries for the vector and hypermultiplet
sectors which contain enough degrees of freedom to allow a scalar
matter field and an inflaton after the truncation.
 The matter field is found in the hypermultiplet
sector where we use the
standard quaternionic geometry
$\frac{\mbox{Sp}(1,1)}{\mbox{Sp}(1)\mbox{Sp}(1)}$ with metric 
\begin{equation}
ds^2 = dh^2 +e^{-2h}(db_1^2 +db_2^2 +db_3^2) 
\end{equation}
and connections
\begin{equation}
\omega^x=\frac{-1}{2}e^{-h}db^x
\end{equation}
If $n_H=1$ the truncation of this is a K\"{a}hler-Hodge manifold which  depends on
only one complex scalar field.  This will be the
matter field. The inflaton is found in the vector sector of the
N=2 theory.  
We take $n_V=2$ so that the special-K\"{a}hler
manifold contains a six component symplectic section.
We choose the prepotential
\begin{equation}
\label{prepot}
F(X)=\frac{-i}{2}((X^0)^2 -(X^1)^2-(X^2)^2)
\end{equation}
which gives the minimal special geometry
\begin{equation}
\label{FandX}
X^{\mathbf{\Lambda}}=\left(\begin{array}{c}
1\\
z_1\\
z_2
\end{array}\right),\;\;\;\;
F_{\mathbf{\Lambda}}=\left(\begin{array}{c}
-iX^0\\
iX^1\\
iX^2
\end{array}\right)=\left(\begin{array}{c}
-i\\
iz_1\\
iz_2
\end{array}\right)
\end{equation}
with $n_C=1$ so that $X={0,1}$ and $\Lambda={2}$ a consistent truncation requires $L^{\Lambda}=0$ (\ref{Lcon}), hence $X^2=z_2=0$.
This leaves the scalar field $z_1$ to be the inflaton which from now
on we write as $z$.  Note that  smaller special-K\"{a}hler
manifolds do not have enough scalar degrees of freedom to
give an inflaton after truncation.  The K\"{a}hler potential (\ref{kahler}) of this
geometry is
\begin{eqnarray}
\label{kahlerz}
K_V&=&-\ln(2(1-z\bar{z}))
\end{eqnarray}
$0\leq|z|<1$.  The vector kinetic matrix (\ref{veckin}) can be
computed  from the prepotential (\ref{prepot}) as
\begin{equation}
\mathcal{N}_{\mathbf{\Lambda}\mathbf{\Sigma}}=\frac{i}{(z^2-1)}\left(\begin{array}{ccc}
1+z^2 & -2z & 0\\
-2z & 1+z^2 & 0\\
0 & 0 & 1-z^2
\end{array}\right)
\end{equation}
The entry needed to compute the D-term of the reduced theory
(\ref{DwithP0}) is
\begin{equation}
\label{imf}
\mathcal{N}_{22}=-i
\end{equation}

Choosing $n_V=2$ gives us the freedom to gauge up to three isometries
in the hypermultiplet sector.  The simplest case is to gauge just one,
and this is the obvious supergravity analogue of the U(1) gauging in the
supersymmetric case.  However in the next section it will be shown that the simplest gauging
does not allow P-term potentials.  In the following sections we consider theories containing one, two and
three Killing vectors in turn, showing whether or not P-term
potentials are allowed and considering the vacuum structure of the
potential and it's suitability for inflation.  To find trajectories
suitable for inflation we look for directions along which the matter
fields are minimised but $z$ is unconstrained.  We shall
call such a direction a valley for $z$.

\subsubsection{Gauging One Isometry}
\label{gauge1}
The simplest gauging is to include only one Killing vector.  We start with a review of how a D-term potential can be constructed. Choose the Killing
vector \cite{Achucarro:2005vz}
\begin{eqnarray}
\label{killing}
 k_{\Lambda} =& & 4b_3 \frac{\partial}{\partial h}  -2(\xi b_2 -2b_1
b_3)\frac{\partial}{\partial b_1} +2(\xi b_1 +2b_2
b_3)\frac{\partial}{\partial b_2}\nonumber\\
& &+2(b_3^2 -e^{2h} +1
-b_1^2-b_2^2)\frac{\partial}{\partial b_3}
\end{eqnarray}
The moment map corresponding to this Killing vector is
\begin{equation}
 P_{\Lambda|N=2} = \left(\begin{array}{c}
-2e^{-h}b_1b_3 -2b_2 +\xi e^{-h}b_2 \\
-2e^{-h}b_2b_3 +2b_1 -\xi e^{-h}b_1\\
-e^{-h}(b_3^2 +1 -b_1^2 -b_2^2) - e^h +\xi \\
\end{array}\right)
\end{equation}
The conditions for truncation (\ref{conDsup}) and (\ref{conkill}) are satisfied on the K\"{a}hler-Hodge
manifold defined by $b_1=b_2=0$.  The moment map reduces to
\begin{equation}
\label{PforD}
 P_{\Lambda|N=1} = \left(\begin{array}{c}
0\\
0\\
\xi-e^h-e^{-h}-e^{-h}b_3^2 \\
\end{array}\right)
\end{equation}
Using (\ref{imf}) the D-term (\ref{DwithP0}) becomes
\begin{equation}
\label{sugraD}
D=\frac{2i(1+\Phi\bar{\Phi})}{\Phi-\bar{\Phi}}-\xi
\end{equation}
where $\Phi=b_3+ie^h$. $D$  contains a constant FI term.

As in the supersymmetric theory it is most straightforward to describe
the P-term potential in N=2 as a rotation from the D-term case.
(\ref{killing}) can be generalised by performing an inverse
SO(3) rotation $R^{-1}_{ij}(\theta,\psi,\phi)$ on the $b_i$ fields.
\begin{eqnarray}
k_{\mathbf{\Lambda}} &=& 4R_{j3}b_j \frac{\partial}{\partial
  h}+\Bigl(4R_{j3}(R_{i1}R_{k1}+R_{i2}R_{k2})b_i b_j
  +2\Bigl((R_{i3}b_i)^2-e^{2h}+1\nonumber\\
& &-(R_{i1}b_i)^2 -
  (R_{i2}b_i)^2\Bigr)R_{k3}+2\tilde{\xi}( R_{i1}R_{k2}-R_{i2}R_{k1})b_i\Bigr)\frac{\partial}{\partial b_k}
\label{kforF}
\end{eqnarray}
with corresponding momentum map
\begin{eqnarray}
P_{x\mathbf{\Lambda}|N=2}& =&-2R_{x3}e^h
-2\epsilon_{xyz}R_{z3}b_y-e^{-h}(2R_{y3}b_yb_x
 -R_{x3}b_yb_y)\nonumber\\
& &-e^{-h}(1-e^{2h})R_{x3}-e^{-h}\tilde{\xi}(R_{y1}R_{x2}-R_{y2}R_{x1})b_y \nonumber\\
& & +\tilde{\xi}
 \epsilon_{xyz}R_{y1}R_{z2}
\label{killingP}
\end{eqnarray}
Note that a generic rotation matrix $R_{ij}$ gives constant terms in all three
components of the moment map.

Truncation to an N=1
theory (\ref{conDsup}) means that either the first two components or
the last component of the moment map must vanish after the truncation.
 Which components vanish depends on whether the index of the
Killing vector lies in the retained vector or retained scalar sector.
As the gauge group is Abelian this means that either the
D-term or the superpotential vanishes.  The two allowed
cases are:
\begin{description}
\item[Vanishing Superpotential.]The truncation conditions  $P_{\Lambda}^1=P_{\Lambda}^2=0$,
  $k_{\Lambda}^t=0$ require
\begin{equation}
\begin{array}{c}
\sin\theta=0\\
b_1=b_2=0
\end{array}
\end{equation}
which is the D-term case considered above with Killing vector
(\ref{killing}) and reduced moment map (\ref{PforD}).

The scalar potential is 
\begin{equation}
\label{Dpot}
V_D=\frac{1}{2}\left(\frac{2i(1+\Phi\bar{\Phi})}{\Phi-\bar{\Phi}}-\xi\right)^2
\end{equation}
The potential is flat for $z$ and so naturally gives rise to
inflation.  We would expect the one loop corrections to this potential
to lift the flat direction so that it is gently sloped.  This may also
give rise to a natural end for inflation as in the hybrid inflationary
scenario discussed in section \ref{rigidinflate}. The supersymmetric minimum of
this potential is at
\begin{equation}
2i(1+\Phi\bar{\Phi})=\xi(\Phi-\bar{\Phi})
\end{equation}
The topology of the vacuum is $S^1$ and cosmic strings can form by the Kibble
mechanism, such strings were studied in \cite{Achucarro:2005vz} and were found to be
BPS solutions.

\item[Vanishing D-term.] The truncation conditions $P_X^3=0$,
  $k_X^s=0$ can only be satisfied if
\begin{equation}
\begin{array}{c}
\cos\theta=0\\
\end{array}
\end{equation}
 so after performing a rotation $R_{ij}(\theta=\pi/2,\psi,\phi)$ on
 the $b_i$ fields the Killing vector (\ref{kforF}) is
\begin{eqnarray}
k_X&=&4(\sin\psi b_1+\cos\psi b_2)\frac{\partial}{\partial
  h}\nonumber\\
&+&\Bigl(2\sin\psi(-e^{2h}+1-b_3^2+b_1^2-b_2^2)+2\cos\psi(\tilde{\xi} b_3
  +2b_1b_2)\Bigr)\frac{\partial}{\partial b_1}\nonumber\\
&+&\Bigl(2\cos\psi(-e^{2h}+1-b_3^2-b_1^2+b_2^2)+2\sin\psi(2b_1b_2-\tilde{\xi}
  b_3)\Bigr)\frac{\partial}{\partial b_2}\nonumber\\
&+&\Bigl(4b_3(\sin\psi b_1+\cos\psi b_2)+2\tilde{\xi}(-b_1
  \cos\psi+b_2\sin\psi)\Bigr)\frac{\partial}{\partial b_3}
\label{Fkilling}
\end{eqnarray}
The choice $\theta=\pi/2$ can be made without loss of generality.  Then the
truncation conditions are 
\begin{equation}
b_1=b_2=0
\end{equation}

The truncated moment map is 
\begin{equation}
\label{FP}
P_{X|N=1}=\left(\begin{array}{c}
\sin\psi\\
\cos\psi\\
0
\end{array}\right)(e^{-h}(b_3^2-1)-e^h+\tilde{\xi})+\left(\begin{array}{c}
\cos\psi\\
-\sin\psi\\
0
\end{array}\right)b_3(2-e^{-h}\tilde{\xi})
\end{equation}
To include $z$, the inflaton field, in the superpotential we choose the
 Killing vector (\ref{Fkilling}) to be $k_1$, then with the K\"{a}hler potential
 (\ref{kahlerz}) and setting $\Phi=b_3+ie^h$ (\ref{Lequation})
 becomes
\begin{equation}
\label{LzPhi}
L=\frac{ize^{i\psi_1}}{\sqrt{2(1-z\bar{z})}}\left(\frac{\Phi^2-1-i\tilde{\xi}\Phi}{\Phi-\bar{\Phi}}\right)
\end{equation}
which is composed of two parts; the total K\"{a}hler potential 
\begin{eqnarray}
K&=&K_V+K_Q\nonumber\\
&=&-\ln(2(1-\bar{z}z))-2\ln(-i(\Phi-\bar{\Phi}))
\end{eqnarray}
and the superpotential 
\begin{equation}
W=ze^{i\psi_1}(\Phi^2-1-i\tilde{\xi}\Phi)
\end{equation}
The reduced
scalar potential, (\ref{potential}), is
\begin{eqnarray}
V_F=&
&\frac{2(1-3z\bar{z})}{1-z\bar{z}}\left|\frac{\Phi^2-1-i\tilde{\xi}\Phi}{\Phi-\bar{\Phi}}\right|^2+\frac{z\bar{z}}{1-z\bar{z}}\left|\frac{2-2\Phi\bar{\Phi}+i\tilde{\xi}(\Phi+\bar{\Phi})}{\Phi-\bar{\Phi}}\right|^2
\label{fullpot}
\end{eqnarray}
which is an F-term potential.  

To see if this potential can give rise to inflation notice that the potential can be extremized in the $\Phi$ directions by setting
$\Phi=i$ ($b_3=h=0$).  This is a minimum if both the following conditions hold:
\begin{eqnarray}
z\bar{z}(6-\tilde{\xi})&>&2-\tilde{\xi}\\
z\bar{z}(3\tilde{\xi}-2)&>&\tilde{\xi}-2
\end{eqnarray}
Along this trajectory the potential is
\begin{equation}
 V(\Phi=i)=\frac{(\tilde{\xi}-2)^2(1-3z\bar{z})}{2(1-z\bar{z})}
\end{equation}
If $\tilde{\xi}\neq2$ the fields are not always confined to this trajectory so for inflation to occur the
initial conditions must be chosen carefully.  Even so the potential
rapidly becomes steeply sloped violating the slow roll conditions.  If
$\tilde{\xi}=2$, then $\Phi=i$
is a flat direction that minimises the potential;  inflation can occur
as $z$ rolls along this valley.

The supersymmetric minimum of the
potential occurs at $\nabla_zL=\nabla_{\Phi}L=0$ which requires 
\begin{equation}
\begin{array}{c}
z=0\\
\Phi^2-1-i\tilde{\xi}\Phi=0
\end{array}
\end{equation}
So the supersymmetric vacuum is unique if $\tilde{\xi}=2$ and no
topological defects form.  If $\tilde{\xi}\neq 2$ then there are two
possible supersymmetric vacuua which may give rise to domain walls.

\end{description}

With a simple gauging it is possible to construct a triplet of
moment maps containing constant terms in N=2
supergravity  (\ref{killingP}).  However subsequent
truncation to N=1 supergravity means that the potential is either an F-term potential (\ref{fullpot}) or a
D-term potential (\ref{Dpot}).  Combinations of the two types which we would
have expected in a P-term model are no longer allowed.

\subsubsection{Gauging Two Isometries}
\label{sec:gauging}
With a simple gauging the supergravity truncation conditions allow us
to have either an F-term potential or a D-term potential but not a
combination of the two.  However by gauging more than one
symmetry it is possible to get FI terms in the D-term and
superpotential of the theory.  
The simplest change we can make to the gauge group is to include two
Killing vectors instead of just one.
We choose $k_1$ to take the form (\ref{Fkilling}) in order to generate an
F-term and $k_2$ to take the form (\ref{killing}) in order to
generate a D-term after truncation.   We relabel $\tilde{\xi}=\xi_1$ and $\xi=\xi_2$ so
that the origin of each FI term is clear.   The Killing
vectors must satisfy
\begin{equation}
[k_{\mathbf{\Delta}},k_{\mathbf{\Sigma}}]=f_{\mathbf{\Delta\Sigma}}^{\mathbf{\Gamma}}k_{\mathbf{\Gamma}}
\end{equation}
where $f_{\mathbf{\Delta\Sigma}}^{\mathbf{\Gamma}}$ are the structure
constants of the gauge group.  This requires
\begin{equation}
\label{xixi}
\xi_1=-\xi_2=\pm 2
\end{equation}
\begin{equation}
f^1_{12}=f^1_{21}=f^2_{12}=f^2_{21}=0
\end{equation}
which gives an Abelian gauge group. 

A consistent truncation can be achieved by setting $b_1=b_2=0$ and $L^2=0$ and reducing the geometry to that considered in the previous section.  Notice that after
truncation only $k_2$ lies in the reduced K\"{a}hler-Hodge manifold,
so the reduced N=1 theory has a U(1) gauging. The reduced potential is
constructed from  (\ref{imf}), (\ref{sugraD}) and (\ref{LzPhi}) 
\begin{eqnarray}
V=&
&\frac{2(1-3z\bar{z})}{1-z\bar{z}}\left|\frac{\Phi^2-1-i\xi_1\Phi}{\Phi-\bar{\Phi}}\right|^2+\frac{z\bar{z}}{1-z\bar{z}}\left|\frac{2-2\Phi\bar{\Phi}+i\xi_1(\Phi+\bar{\Phi})}{\Phi-\bar{\Phi}}\right|^2
\nonumber\\
&+&\frac{1}{2}\left(\frac{2i(1+\Phi\bar{\Phi})}{\Phi-\bar{\Phi}}-\xi_2\right)^2
\end{eqnarray}
where the FI terms are constrained by  (\ref{xixi}).

To look for inflation in this potential notice that $\Phi=i$ is an extremum in the $\Phi$ directions.  Along this direction
\begin{equation}
V(\Phi=i)=\left\{\begin{array}{lll}
\frac{8(1-3z\bar{z})}{1-z\bar{z}} & \mbox{ if }\xi_1=-2, &  \xi_2=2\\
8 & \mbox{ if }\xi_1=2, & \xi_2=-2
\end{array}\right.
\end{equation}
If $\xi_1=-2$, $\xi_2=2$ then $\Phi=i$ is only a minimum if $z\bar{z}<1/2$
and the potential rapidly becomes steeply sloped once the fields move
away from the maximum at $z=0$.  It seems unlikely that sufficient
efolds of inflation could occur in such a potential.
If $\xi_1=2$, $\xi_2=-2$ then $\Phi=i$ is a global minimum and
inflation would occur as $z$ rolls along this flat direction.

A supersymmetric minimum requires $\nabla_{\Phi}L=\nabla_zL=D=0$
but because $\xi_1$ and $\xi_2$ are not equal there is no
solution to these equations.  There is no supersymmetric minimum in
this system.

\subsubsection{Gauging Three Isometries}
\label{gauge3}
The simple geometry we have chosen allows us to gauge up to three
isometries.  Including three Killing vectors of the form we have been
considering gives a P-term potential which behaves differently to that
of the previous section.  We take one Killing
vector $k_2$ of the form (\ref{killing}) to generate the D-term and two linearly independent
vectors $k_0$, $k_1$ each of the form (\ref{Fkilling}) to generate the
F-term.  For $k_0$, $k_1$ to be linearly independent we must have
$\sin(\psi_0-\psi_1)\neq0$, where $\psi_i$ is the Euler angle appearing in the expression for $k_i$ (\ref{Fkilling}).  For these Killing vectors to generate a group,
we must impose
\begin{equation}
\xi_0=\xi_1=\xi_2=2
\end{equation}
and the structure constants are
\begin{eqnarray}
f^0_{01}&=&f^1_{01}=f^2_{21}=f^2_{20}=0\\
f^1_{20}&=&-f^0_{21}=\frac{8}{\sin(\psi_1-\psi_0)}\\
f^1_{21}&=&-f^0_{20}=\frac{8\cos(\psi_1-\psi_0)}{\sin(\psi_1-\psi_0)}\\
f^2_{01}&=&8\sin(\psi_1-\psi_0)
\end{eqnarray}
and their permutations.

A consistent truncation requires $b_1=b_2=0$ and $L^2=0$.  Notice that
after the truncation only $k_2$ lies in the reduced K\"{a}hler-Hodge
manifold, so the N=1 gauge group is U(1).  However, as $P^0_{\Sigma}$ no longer vanishes,  
the potential contains relics of the original N=2 non-Abelian gauge group. 
The moment maps are computed from $k_0$,
$k_1$, $k_2$ as before and
\begin{equation}
P^0_2=\frac{8i\sin(\psi_1-\psi_0)(z-\bar{z})}{1-z\bar{z}}
\end{equation}
which gives
\begin{equation}
L=\frac{i(e^{i\psi_0}+ze^{i\psi_1})}{\sqrt{2(1-z\bar{z})}}\left(\frac{\Phi^2-1-2i\Phi}{\Phi-\bar{\Phi}}\right)
\end{equation}
and
\begin{equation}
D=\frac{2i(1+\Phi\bar{\Phi})}{\Phi-\bar{\Phi}}-2+\frac{8i\sin(\psi_1-\psi_0)(z-\bar{z})}{1-z\bar{z}}
\end{equation}
The reduced scalar potential is
\begin{eqnarray}
V=&
&\frac{8|1+ze^{i(\psi_1-\psi_0)}|^2}{|1-z\bar{z}|}\left(\frac{i(1+\Phi\bar{\Phi})-(\Phi-\bar{\Phi})}{\Phi-\bar{\Phi}}\right)\\
& &+\frac{1}{2}\left(\frac{2i(1+\Phi\bar{\Phi})}{\Phi-\bar{\Phi}}-2+\frac{8i\sin(\psi_1-\psi_0)(z-\bar{z})}{1-z\bar{z}}\right)^2
\end{eqnarray}

$\Phi=i$ extremizes the potential in the $\Phi$ directions and is a minimum if 
\begin{equation}
z\bar{z}+i(2\sin(\psi_1-\psi_0)+1)(z-\bar{z})+1>0
\end{equation}
The potential along this direction is
\begin{equation}
V(\Phi=i)=\frac{32\sin^2(\psi_1-\psi_0)|z-\bar{z}|^2}{(1-z\bar{z})^2}
\end{equation}
We can tune the slope of the potential so that inflation can occur in
this valley.  the fields will eventually settle into the minimum at $z=0$.
The conditions for a supersymmetric minimum $\nabla_zL=0$,
$\nabla_{\Phi}L=0$ and $D=0$ 
 are satisfied when 
\begin{equation}
\Phi=i,\;\;\;\;\;z=\bar{z}
\end{equation}
so the minimum reached at the end of inflation is supersymmetric.

\section{Conclusion}
\label{conc}
The study of P-term scalar potentials in supergravity is interesting
because the P-term model is the effective theory of the D3/D7 brane inflation
model, one of the most
promising models of string theory inflation, .  P-term models have been studied in supersymmetry and in rigid
supergravity but there were thought to be problems involved in lifting
the model to supergravity.  Previously the P-term model has been
studied in
the rigid limit of supergravity where all P-term potentials can inflate and with some tuning of the
parameters can produce sufficient efolds of inflation and density
perturbations of the correct size to agree with observations.  Cosmic
strings are formed at the end of inflation.

The major obstacle to constructing a P-term theory in supergravity was
that it was not known how to include FI terms in N=2 supergravity.  We
have shown that FI terms can be produced in both the superpotential
and the D-term of the theory by considering the truncation to an N=1
theory.  If only one isometry is gauged this truncation restricts the
form of the N=1 scalar potential so that it contains either F-terms
or D-terms but not both.  It is not possible to construct a P-term
model with a simple gauging by this method.  If more than one isometry
is gauged the potential can include F-terms and D-terms both
containing FI constants.  Although the gauge group of the N=2 theory
is now more complicated the gauge group of the reduced N=1 theory is
always the simple Abelian case where only one isometry is gauged.
Therefore we believe these models are the supergravity analogues of
the well known supersymmetry P-term potentials.  

One of the nice properties of the supersymmetric theory was that by
rotating the vector of FI terms it was possible to move from a D-term
potential to an F-term one, and the magnitude of the FI vector was undetermined.  This is no longer true in supergravity,
where after the truncation to an N=1 theory the position of the FI
terms is fixed.  If we gauge more than one symmetry the size of the FI
terms is also fixed.  In each of the potentials we have considered we
have demonstrated a direction
which could give rise to inflation.  It is a subject for further
research whether these potentials give rise to sufficient efolds of
inflation, and whether they match cosmological observations.  This is
in progress.  In these classes of models cosmic strings are not formed
at the end of inflation.  This could alleviate the parameter
constraints on such models.

\section*{Acknowledgements}
We would like to thank P. Brax and M. Esole for useful discussions,
and A. Achucarro, A. Van Proeyen, R. Kallosh and  K. Sousa for
clarifying comments.  This work is supported in part by PPARC.  We
thank CEA Saclay, the Galileo Galilei Institute for Theoretical
Physics and the Leiden Institute of Physics for their
hospitality and the INFN and the European Science Foundation (ESF) for partial support during the completion of this
work.

\appendix
\section{Euler Angles}
\label{angles}
We use the following parameterisations for an SO(3) rotation in terms
of Cayley-Klein parameters
\begin{equation}
R=\left(\begin{array}{ccc}
\frac{1}{2}(a^2-b^{*2}+a^{*2}-b^2) &
\frac{i}{2}(b^{*2}-a^2+a^{*2}-b^2) & -ab-a^*b^*\\
\frac{i}{2}(a^2+b^{*2}-a^{*2}-b^2) &
\frac{1}{2}(b^{*2}+a^2+a^{*2}+b^2) & -i(ab-a^*b^*)\\
ba^*+ab^* & i(ba^*-ab^*) & aa^*-bb^*
\end{array}\right)
\end{equation}
which are defined in terms of Euler angles as
\begin{equation}
a=e^{i(\psi+\phi)/2}\cos\frac{\theta}{2}
\end{equation}
\begin{equation}
b=ie^{i(\psi-\phi)/2}\sin\frac{\theta}{2}
\end{equation}
The SU(2) rotation associated with this SO(3) rotation is 
\begin{equation}
U=\left(\begin{array}{cc}
a & b\\
-b^* & a^*
\end{array}\right)
\end{equation}

\section{The Effective Potential}
\label{fulleffpot}
The effective potential in Supergravity is given by (\ref{formofpot}),
an explicit calculation of this with the K\"{a}hler potential
(\ref{mink}), superpotential (\ref{sup}) and D-term (\ref{D})
\begin{eqnarray}
V=&
&\frac{g^2}{2}e^K\left\{|\phi_+\phi_-|^2\left(1+\frac{|\phi_0|^4}{M_P^4}\right)+|\phi_0\phi_-|^2\left(1+\frac{|\phi_+|^4}{M_P^4}\right)\right.\nonumber\\
&
&+|\phi_0\phi_+|^2\left(1+\frac{|\phi_-|^4}{M_P^4}\right)+\frac{3|\phi_0\phi_+\phi_-|^2}{M_P^2}\nonumber\\
& &-\sin\theta\xi(e^{i\psi}\phi_+\phi_-+e^{-i\psi}\bar{\phi}_+\bar{\phi}_-)\nonumber\\
&
&\;\;\;\;\times\left(1+\frac{|\phi_0|^2}{M_P^2}+\frac{|\phi_0|^2}{M_P^4}(|\phi_0|^2+|\phi_+|^2+|\phi_-|^2)\right)\nonumber\\
& &+(\sin\theta\xi)^2\nonumber\\
& &\;\;\;\;\left.\times\left(1-\frac{|\phi_0|^2}{M_P^2}+\frac{|\phi_0|^2}{M_P^4}(|\phi_0|^2+|\phi_+|^2+|\phi_-|^2)\right)\right\}\nonumber\\
&+&\frac{g^2}{8}(|\phi_+|^2-|\phi_-|^2-\cos\theta\xi)^2
\end{eqnarray}

\bibliography{reportbib}
\end{document}